 \documentclass[referee]{aa} 
\usepackage{graphicx}
\usepackage{txfonts}

\usepackage{natbib,twoopt}
 \usepackage[breaklinks=true]{hyperref} 

 \bibpunct{(}{)}{;}{a}{}{,}    
 \newcommandtwoopt{\citeads}[3][][]{\href{http://adsabs.harvard.edu/abs/#3}%
                                        {\citealp[#1][#2]{#3}}}
 \newcommandtwoopt{\citepads}[3][][]{\href{http://adsabs.harvard.edu/abs/#3}%
                                        {\citep[#1][#2]{#3}}}
 \newcommandtwoopt{\citetads}[3][][]{\href{http://adsabs.harvard.edu/abs/#3}%
                                        {\citet[#1][#2]{#3}}}
 \newcommandtwoopt{\citeyearads}[3][][]%
   {\href{http://adsabs.harvard.edu/abs/#3}{\citeyear[#1][#2]{#3}}}

\usepackage{txfonts}

\begin{document}
   \title{The GTC exoplanet transit spectroscopy survey. IV.: No asymmetries in the transit of Corot-29b}

 
   \author{E. Pall\'e\inst{1,2}, G. Chen\inst{1,2,3}, R. Alonso\inst{1,2}, G. Nowak\inst{1,2}, H. Deeg\inst{1,2}, J. Cabrera\inst{4},  F. Murgas\inst{5,6}, H. Parviainen\inst{7}, L. Nortmann\inst{8}, S. Hoyer\inst{1,2}, J. Prieto-Arranz\inst{1,2}, D. Nespral\inst{1,2}, A. Cabrera Lavers\inst{1,2}, N. Iro\inst{9}
          }

   \institute{Instituto de Astrof\'isica de Canarias (IAC), V\'ia L\'actea s/n 38205, La Laguna, Spain
         \and
             Departamento de Astrof\'isica, Universidad de La Laguna, Spain
\and
        Key Laboratory of Planetary Sciences, Purple Mountain Observatory, Chinese Academy of Sciences, Nanjing 210008, China
\and
Institute of Planetary Research, German Aerospace Center, Rutherfordstrasse 2, D-12489 Berlin, Germany
\and
Univ. Grenoble Alpes, IPAG, F-38000 Grenoble, France
\and
CNRS, IPAG, F-38000 Grenoble, France
\and
Sub-department of Astrophysics, Department of Physics, University of Oxford, Oxford, OX1 3RH, UK
\and
Institut f\"ur Astrophysik, Georg-August-Universit\"at G\"ottingen, Friedrich-Hund-Platz 1, D-37077 G\"ottingen, Germany 
\and
Theoretical Meteorology group, Klimacampus, University of Hamburg, Grindelberg 5, 20144 Hamburg, Germany\\
              \email{epalle@iac.es}
             }

   \date{Received December 2, 2015; accepted January 30, 2016}

 
  \abstract
   { The launch of the exoplanet space missions obtaining exquisite photometry from space has resulted in the discovery of thousands of planetary systems with very different physical properties and architectures. Among them,  the exoplanet CoRoT-29b was identified in the light curves the mission obtained in summer 2011, and presented an asymmetric transit light curve, which was tentatively explained via the effects of gravity darkening.}
   {Transits of CoRoT-29b are measured with precision photometry, to characterize the reported asymmetry in their transit shape.}
   {Using the OSIRIS spectrograph at the 10-m GTC telescope, we perform spectro-photometric differential observations, which allow us to both calculate a high-accuracy photometric light curve, and a study of the color-dependence of the transit.}
   {After careful data analysis, we find that the previously reported asymmetry is not present in either of two transits, observed in July 2014 and July 2015 with high photometric precisions of 300ppm over 5 minutes. Due to the relative faintness of the star, we do not reach the precision necessary to perform transmission spectroscopy of its atmosphere, but we see no signs of color-dependency of the transit depth or duration. }
   {We conclude that the previously reported asymmetry may have been a time-dependent phenomenon, which did not occur in more recent epochs. Alternatively, instrumental effects in the discovery data may need to be reconsidered.}

   \keywords{giant planet formation --
                $\kappa$-mechanism --
                stability of gas spheres
               }


\authorrunning{Palle et al}

\titlerunning{CoRoT-29b no asymmetries}

   \maketitle
%

\section{Introduction}

The launches of the exoplanet space missions CoRoT (\citeads{2003AdSpR..31..345B}) and Kepler (\citeads{2003SPIE.4854..129B}), returning exquisite photometry from space of thousands of stars, represented a tremendous milestone in exoplanet research. These missions have discovered hundreds of confirmed exoplanets, from hot Jupiters to rocky planets (\citeads{2009A&A...506..287L}, \citeads{2011ApJ...728..117B}), some of them possibly habitable (\citeads{2013Sci...340..587B}, \citeads{2015AJ....150...56J}). 

Among the possibilities that such exquisite photometry allow is the exploration of the interaction between stars and planets.
Gravity darkening is produced by the oblateness of a star due to fast rotation or the gravitational pull of a close companion. Indeed, gravity darkening, which takes into account the relationship between surface brightness and local effective gravity, is necessary to explain the light curves of many binary systems (\citeads{2012A&A...547A..32E}).

In the context of exoplanets, two transiting hot Jupiter systems monitored with the Kepler spacecraft, Kepler-13Ab  (\citeads{2012A&A...541A..56M}) and HAT-P-7b (\citeads{2008ApJ...680.1450P}) have shown evidences of gravity darkening in their asymmetric transit light curves (\citeads{2011ApJS..197...10B}; \citeads{2013ApJ...764L..22M}; \citeads{2015ApJ...805...28M}). 

More recently, \citetads{2015A&A...579A..36C} reported the discovery of CoRoT-29b, a Jupiter-sized planet ($M = 0.85\pm0.20 M_{Jup}$;  $R = 0.90\pm0.16 R_{Jup}$) orbiting an oblate star with an orbital period of $\approx2.85$ days. Because of the asymmetric shape of its transit light curve, observed both in the CoRoT data and subsequent ground-based followup, transit fitting models had to include the possibility that the stellar surface was not strictly spherical. Thus gravity darkening effects were called upon for a possible explanation, which the authors themselves did not consider definitive. 

Still, CoRoT-29 is a relatively faint star ($m_V =15.56$) and some doubts remained about the significance of the asymmetry detection. In such cases, the use of large collecting area ground-based telescopes may reach the sensitivity that space facilities cannot. Here, we have used the 10-m GTC telescope to investigate the transit light curve of CoRoT-29b in search for gravity darkening signatures.

\begin{figure}
\centering
\includegraphics[width=0.485\hsize]{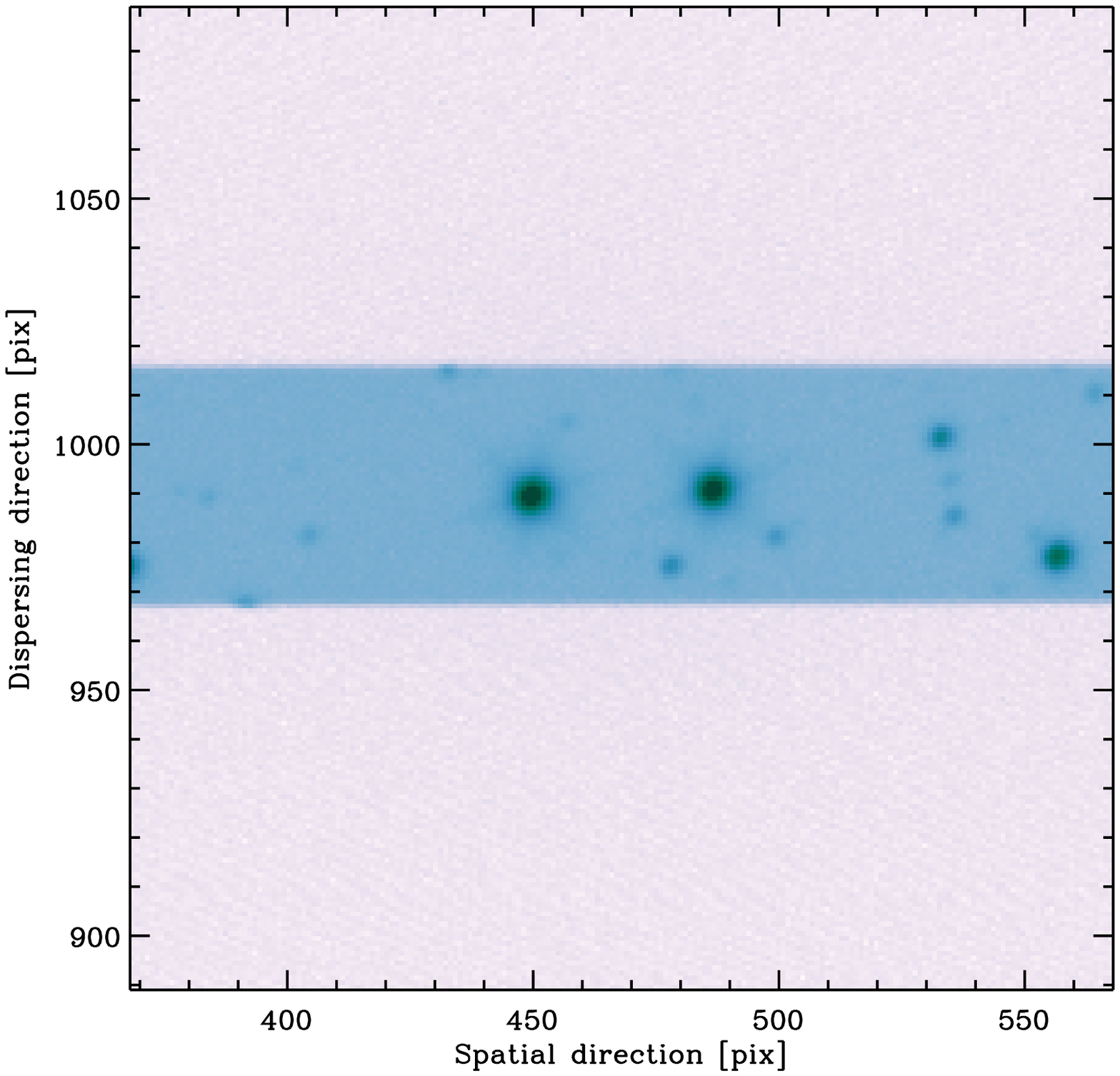}
\includegraphics[width=0.485\hsize]{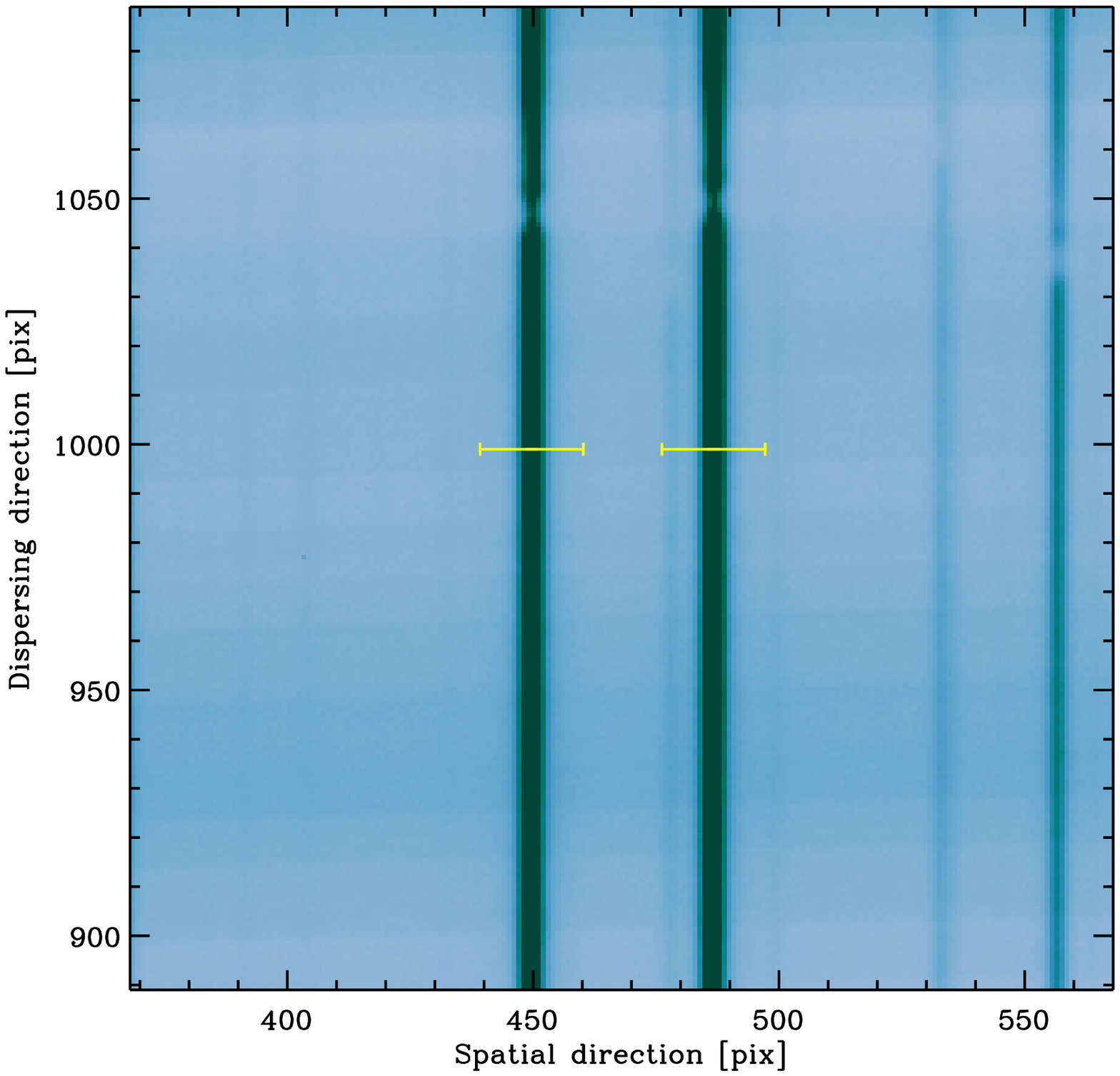}
\label{fig1}
\caption{Slit image (left) and its corresponding dispersed 2D spectra (right). The left of the two bright stars is CoRoT-29. The yellow lines mark the best aperture size that minimizes the light-curve standard deviation. The aperture of the reference star includes the flux of the faint close-companion stars.}
\end{figure}

\begin{table*}
\caption{Observing logs of the two transit datasets, including date, times, observing setup and atmospheric conditions.  The rms values of the observations refer to the point-to-point-rms of the white light curve, when both transits are jointly fitted. Due to different CCD1 windowing sizes, the observing cadences of the two observations are slightly different: 317s and 323s per exposure+overhead. The rms of these light curves are 208ppm and 240ppm, respectively.}
\label{Table1}
\centering
\begin{tabular}{l c c cc c c r c}
\hline\hline
Object 	& Obs. date  & Time 	&  Grating & Slit	&Exposure 	&FWHM	& Airmass & rms       	    \\
 		&(UT)	    &	(UT)             & 		&              & (s)	                 &(")	       & &	(ppm)                                                      \\
\hline
CoRoT-29	&2014/7/31 	& 21:29--03:17    & $R1000R$ 	& 12" & 300   	& 0.74--2.11	&  1.19--2.11& 292 	\\
CoRoT-29	&2015/7/8 	& 00:16--04:40    & $R1000R$ 	& 12" & 300   	& 0.68-2.23	&  1.08--1.98& 311	\\
\hline\hline
\end{tabular}
\end{table*}

\section{Observations and data reduction}

Two full transits of CoRoT-29b were observed with OSIRIS (Optical System for Imaging and low-Intermediate-Resolution Integrated Spectroscopy; \citeads{2012SPIE.8446E..4TS}) spectrograph at the 10-m Gran Telescopio Canarias (GTC). The observed transits epochs were 31 July 2014 and 8 July 2015.

Observations were taken in the 2x2 binning mode, a readout speed of 200 kHz, and a gain of 0.95 e-/ADU, a combination that resulted in a readout noise of 4.5 e-. The R1000R grating was used and a custom-made 12"-wide slit was used. This long-slit spectroscopy configuration covers the spectral range of 520-1040 nm, although wavelengths redder than 910 $nm$ are affected by fringing \citepads{2014A&A...563A..41M} and were not used in this work. 

As illustrated in Figure~\ref{fig1}, both CoRoT-29 and the comparison star (2MASS-18353616+062854), separated in the sky 9.7 arcsec, were located on OSIRIS CCD1, close to the telescope optical center, which reduces considerably all telescope systematics (Chen et al. ; Nortmann et al.; submitted). The position angle of the reference star with respect to the target (measured East of North) was -33.98 degrees. A detailed log of the observations can be found in Table~\ref{Table1}. 

\begin{figure}
\centering
\includegraphics[width=0.95\hsize]{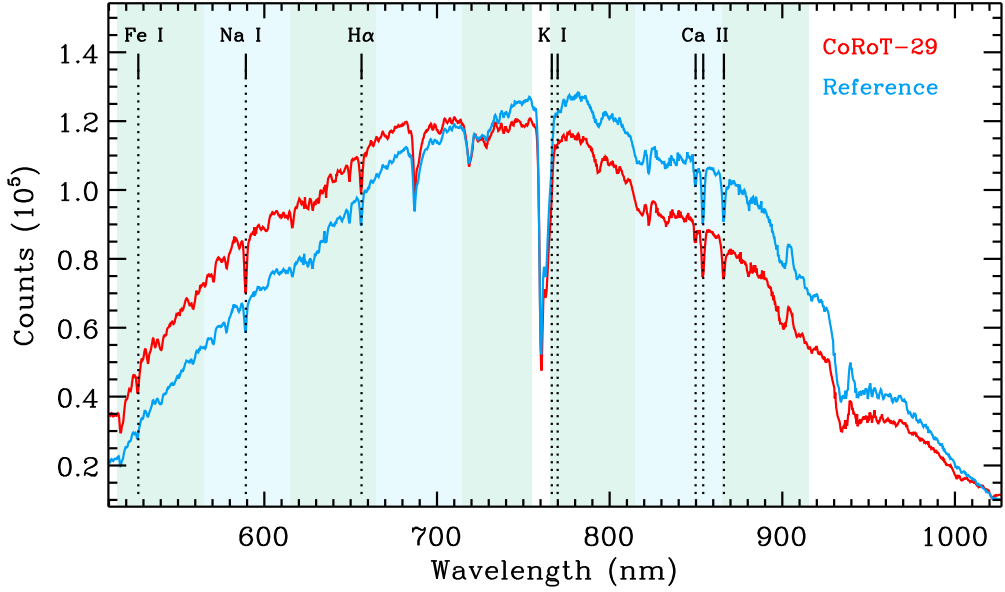}
\label{fig2}
\caption{Extracted spectra of CoRoT-29  (in red), and the reference star (in blue). The spectra were averaged over all out-of-transit observations. The color shaded areas indicate the pass-bands that are used to generate the spectro-photometric light curves. Note that the spectral type of the comparison is a bit redder than CoRoT-29, but has similar spectral features. Some spectral lines, common in stellar and planetary atmospheres, are also marked. The spectra are not corrected for instrumental response.}
\end{figure}

Data were reduced using standard procedures, and include bias substraction, flat fielding, and corrections for spatial distortion and wavelength shifts to bring all the spectra to a reference. This methodology is described in detail in \citetads{2014A&A...563A..41M} and (Chen et al. ; submitted). Spectra extraction was performed using IRAF\footnote{IRAF is distributed by the National Optical Astronomy Observatory, which is operated by the Association of Universities for Research in Astronomy (AURA) under cooperative agreement with the National Science Foundation.}. While Corot-29b has no close companion, our comparison star has two nearby faint companions which were completely included in the extraction aperture.  The optimal aperture diameters were 22 and 21 pixels for the first and second night, respectively, which were retrieved using the built-in {\it optimal extraction} procedure in IRAF. 
Figure~\ref{fig2} shows the extracted spectrum of both CoRoT-29 and the slightly redder reference star.

For each image we calculate the drifts in spatial and also in dispersion direction. We monitored the drift in spatial direction by fitting a Gaussian function to the stellar profile tracing the position of the peaks. The FWHM of the fitted Gaussian profile in the spatial direction is used to monitor the seeing variations. The drift in the dispersion direction was determined by fitting a Gaussian function to the absorption lines (e.g. $H_\alpha$).

\begin{table*}
\caption{CoRoT-29b refined planet parameters.}
\label{Table2}
\centering
\begin{tabular}{cccccc }
   \hline\hline\noalign{\smallskip}
   Parameter & \multicolumn{2}{c}{Individual analysis} & \multicolumn{2}{c}{Joint analysis} & Cabrera et al. (2014)\\\cline{2-3}\cline{4-5}\noalign{\smallskip}
             & 31--Jul--2014 & 08--Jul--2015 & 31--Jul--2014 & 08-Jul-2015 & \\\noalign{\smallskip}
   \hline\noalign{\smallskip}
   $P$ [days]\tablefootmark{a} & \multicolumn{4}{c}{2.8505615 $^{+0.0000072}_{-0.0000072}$} & 2.850570 $^{+0.000006}_{-0.000006}$ \\\noalign{\smallskip}
   $T_0$\tablefootmark{b}    & \multicolumn{4}{c}{6870.53809 $^{+0.00075}_{-0.00075}$} & 5753.115 $^{+0.001}_{-0.001}$ \\\noalign{\smallskip}
   \hline\noalign{\smallskip}
   $R_{\rm p}/R_\star$ & 0.1104 $^{+0.0016}_{-0.0019}$ & 0.1118 $^{+0.0028}_{-0.0029}$ & \multicolumn{2}{c}{0.1105 $^{+0.0013}_{-0.0014}$} & 0.1028 $^{+0.0043}_{-0.0043}$ \\\noalign{\smallskip}
   $T_{\rm mid}$\tablefootmark{b} & 6870.53863 $^{+0.00025}_{-0.00025}$ & 7212.60480 $^{+0.00032}_{-0.00032}$ & 6870.53864 $^{+0.00023}_{-0.00024}$ & 7212.60482 $^{+0.00030}_{-0.00030}$ & ... \\\noalign{\smallskip}
   $i$ [deg] & 85.77 $^{+0.47}_{-0.41}$ & 85.78 $^{+0.54}_{-0.48}$ & \multicolumn{2}{c}{85.80 $^{+0.32}_{-0.30}$} & 87.3 $^{+2.7}_{-2.7}$ \\\noalign{\smallskip}
   $a/R_\star$ & 8.52 $^{+0.34}_{-0.30}$ & 8.47 $^{+0.39}_{-0.33}$ & \multicolumn{2}{c}{8.53 $^{+0.23}_{-0.22}$} & 9.22 $^{+0.19}_{-0.19}$ \\\noalign{\smallskip}
   $u_1$ & 0.442 $^{+0.069}_{-0.071}$ & 0.476 $^{+0.080}_{-0.082}$ & \multicolumn{2}{c}{0.455 $^{+0.051}_{-0.052}$} & $0.60\pm0.15$ \\\noalign{\smallskip}
   $u_2$ & 0.220 $^{+0.091}_{-0.088}$ & 0.225 $^{+0.092}_{-0.091}$ & \multicolumn{2}{c}{0.220 $^{+0.064}_{-0.064}$} & $0.02\pm0.15$\\\noalign{\smallskip}
   $c_0$\tablefootmark{c} & 1.00107 $^{+0.00021}_{-0.00020}$ & 1.00518 $^{+0.00051}_{-0.00049}$ & 1.00112 $^{+0.00021}_{-0.00018}$ & 1.00488 $^{+0.00031}_{-0.00030}$ & ... \\\noalign{\smallskip}
   $c_1$\tablefootmark{c} & -0.000045 $^{+0.000061}_{-0.000062}$ & -0.00242 $^{+0.00012}_{-0.00012}$ & -0.000045 $^{+0.000060}_{-0.000060}$ & -0.00244 $^{+0.00012}_{-0.00012}$ & ... \\\noalign{\smallskip}
   $c_2$\tablefootmark{c} & -0.0133 $^{+0.0016}_{-0.0017}$ & -0.0475 $^{+0.0034}_{-0.0036}$ & -0.0134 $^{+0.0016}_{-0.0017}$ & -0.0471 $^{+0.0032}_{-0.0034}$ & ... \\\noalign{\smallskip}
   $c_3$\tablefootmark{c} & -0.193 $^{+0.020}_{-0.020}$ & -0.642 $^{+0.086}_{-0.087}$ & -0.198 $^{+0.019}_{-0.020}$ & -0.592 $^{+0.057}_{-0.058}$ & ... \\\noalign{\smallskip}
\hline\hline
\end{tabular}
\tablefoot{
\tablefoottext{a}{Value Fixed in MCMC Analysis}
\tablefoottext{b}{For both $T_0$ and $T_{\rm mid}$a constant value of 2450000 has been subtracted. Our times here are $\mathrm{BJD}_\mathrm{TDB}$, while Cabrera's is $\mathrm{HJD  (UTC)}$.}
\tablefoottext{c}{Coefficients in the detrending function: $B=c_0+c_1s_y+c_2t+c_3t^2$, where $s_y$ is the FWHM of spatial PSF.} 
}
\end{table*}

\section{Transit white and color light curves}

Transit light curves were obtained by integrating the fluxes of each spectrum of the time series and then taking the ratio between the integrated flux of the target and the reference star. In the case of the white light curve, the flux was integrated between 515 and 915 nm. 

To study the possible color signature of CoRoT-29b, we defined eight broad filters, whose spectral coverage is illustrated in Figure~\ref{fig2}. Redwards of 915 nm we do not use the data as they might be affected by fringing. The region around the telluric oxygen band at 760 nm is also discarded due to its poor signal-to-noise ratio \citepads{2016A&A...585A.114P}.

 \subsection{Light curve fitting}

To derive the transit parameters, we fit the flux ratio between the target and comparison star light curves with an analytic transit model, simultaneously multiplied by a parametric decorrelation baseline model, of the form:

$F_{\rm mod}=\mathcal{T}(p_i)\times\mathcal{B}(c_j)$.

where the analytic transit model $\mathcal{T}(p_i)$ is taken from \citet{2002ApJ...580L.171M}, and is described by the planet's position relative to the star, assuming an opaque, dark sphere eclipsing a spherical star. The best parametric decorrelation baseline model, $B (c_j)=c_0+c_1s_y+c_2t+c_3t^2$, was selected among several others using the BIC evaluation criteria \citepads{1978A&A....65..345S}, and consists of a polynomial combination of various state vectors, including the FWHM in the spatial direction $s_y$ and time $t$. To determine the best-fitting model parameters and their associated uncertainties, we use a modified version of the Transit Analysis Package \citepads{2012AdAst2012E..30G} to perform a Markov Chain Monte Carlo (MCMC) analysis.

The free parameters in our light-curve model were the mid-transit time $T_{\rm mid}$, the orbital inclination $i$, the scaled semi-major axis $a/R_\star$, the planet-to-star radius ratio $R_{\rm p}/R_\star$, two limb-darkening coefficients ($u_1$, $u_2$), and four baseline coefficients ($c_0$, $c_1$, $c_2$, $c_3$). The orbital period $P$ was first fixed to the literature value and then refined with the newly fitted mid-transit times. The orbital eccentricity $e$ and argument of periastron $w$ were fixed to zero. The transit model uses a quadratic limb-darkening law, with the derived coefficients treated as Gaussian priors in the fitting of light curves. The full methodology employed in the model fitting is extensively detailed in Chen et al. (submitted).

In Table~\ref{Table2}, the refined planet parameters for CoRot-29b are given. The orbital period $P$ and the zero epoch $T_0$ are calculated by fitting a linear function to our two transit epochs together with literature epochs. The rest of parameters are calculated by individually fitting the light curve of each epoch, but also by jointly fitting both transits together. These values are compared to those in the discovery paper by \citetads{2015A&A...579A..36C}.

\begin{figure}
\centering
\includegraphics[width=0.95\hsize]{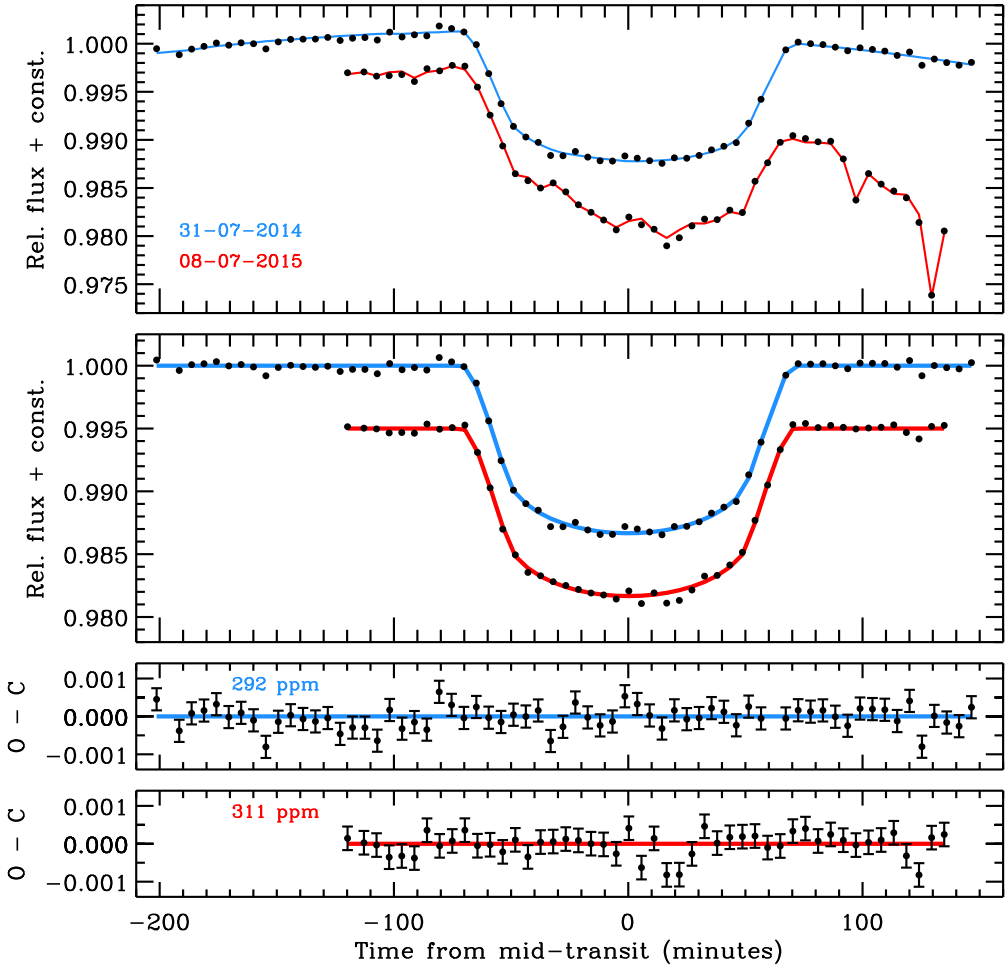}
\label{fig3}
\caption{White light curve of CoRoT-29b. It is a summation of all the color-shaded passbands as shown in Fig.~2. From top to bottom are: (1) normalized flux ratio between CoRoT-29 and its reference star, together with the best-fitting model; (2) white light curve after removing the modeled systematics, for display purpose; (3, 4) best-fitting light-curve residuals.}
\end{figure}

\begin{figure}
\centering
\includegraphics[width=0.45\hsize]{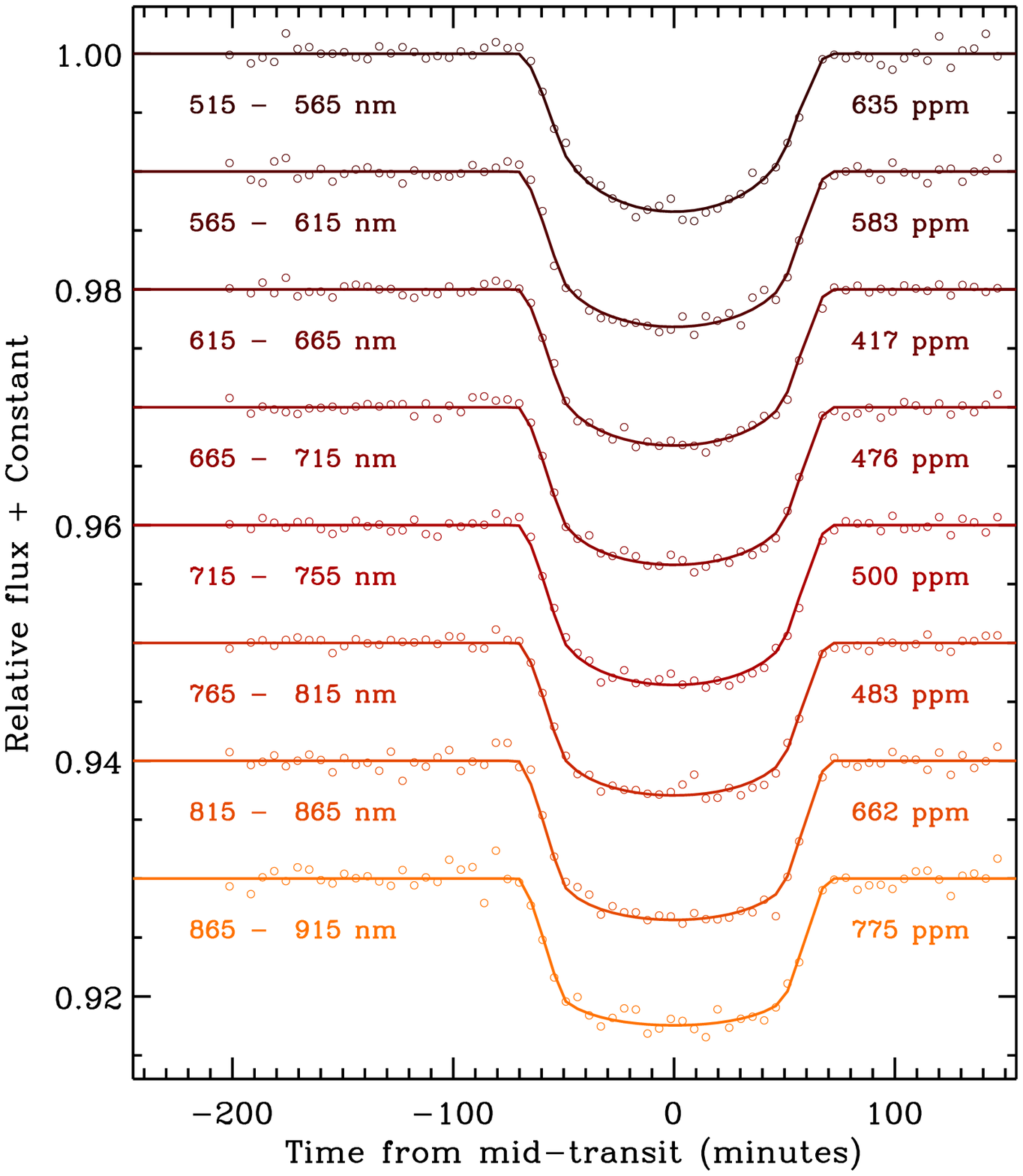}
\includegraphics[width=0.45\hsize]{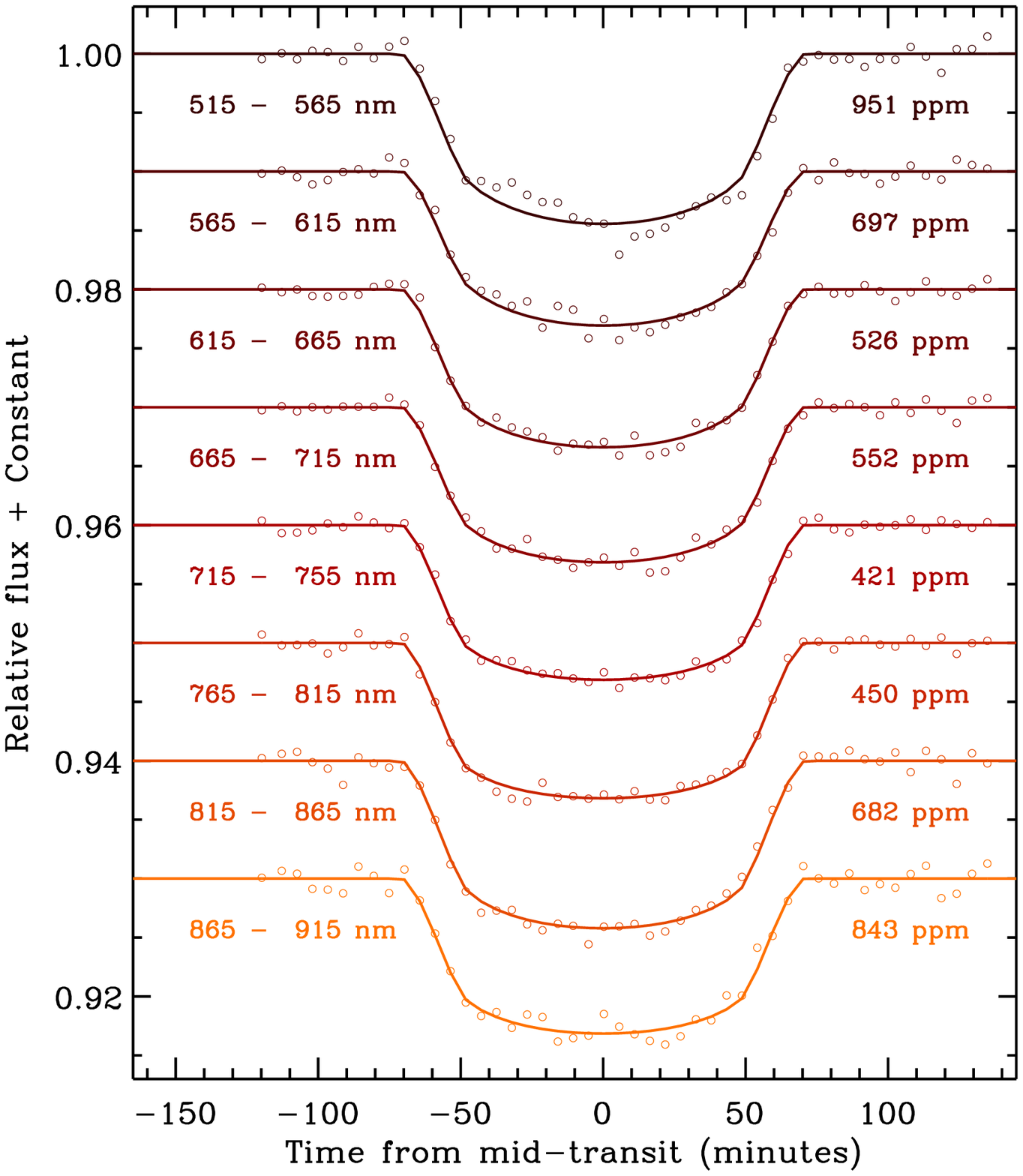}
\label{fig4}
\caption{Color light curves of CoRoT-29b. Left and right panels show the data from July 31st, 2014 and July 8th 2015, respectively. The light curves are shown after removing the modeled systematics, and overplotted are the best-fitting light-curve models.}
\end{figure}

\begin{figure}
\centering
\includegraphics[width=0.8\hsize]{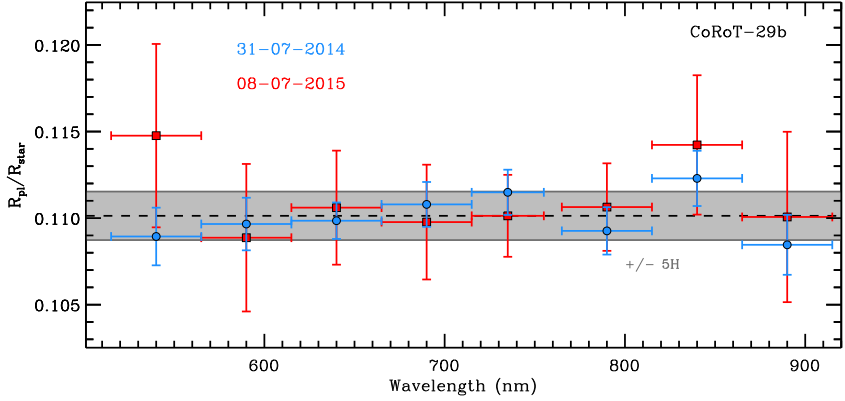}
\label{fig5}
\caption{The transmission spectrum of CoRoT-29b. The dashed line indicates the best-fitting $R_{\rm p}/R_\star$ derived from the white light curve. Blue and red colors indicate the two observing epochs. The gray-shaded area marks the $\pm 5H$ scale height. Within $1\sigma$ error bars the spectrum is flat. }
\end{figure}

\section{Discussion and Conclusions}

Figure~3 presents the white light curve for the two datasets. In the top panel the raw flux ratios between CoRoT-29 and the comparison star are shown, where effects of airmass and seeing are still present. In the middle panel of the same figure, the light curves have been normalized and all systematic effects, except the analytic transit model, have been removed.  Overplotted to the data are our best fits to each of the transits. It is remarkable that even at the faint magnitude of the CoRoT-29 ($m_v = 15.56$), using GTC we are able to reach precisions down to $\approx$300 ppm in 5 minutes exposure time; this is a factor $\approx2.2$ above photon noise. The residuals of the transit fit do not show any clearly identifiable features attributable to red noise, or an unaccounted for planetary signal. 

To be sure that the decorrelation process is not removing real asymmetries in the transit curve, we have generated mock light curves introducing an asymmetry in the form of a stellar spot (not shown). In that case using the same fitting methodology and a symmetric transit, the asymmetry is recovered, and only when using an asymetric transit profile, taking into accout the stellar spot effects, are the residuals of the fit similar to those in Figure~\ref{fig3}.

To further explore possible transit anomalies, we plot in Figure~4 the color light curves for each of our spectral bins for the two observed epochs. It can be seen that the symmetry of the transit is maintained over all spectral bins and in both epochs. The only exception would be on the bluest light curves from July 8th 2015, where deviations from the transit model are larger. However, these anomalies are not only of the opposite sign to those reported in the original discovery paper by \citetads{2015A&A...579A..36C}, but are more likely due to low signal to noise ratios in this wavelength bin. 

While for such a faint star the signal-to-noise ratio is far too low to perform transmission spectroscopy to explore the atmosphere of CoRot-29b, it is still useful to look at the transit parameters as a function of wavelength. In Figure~5 the retrieved transit depth as a function of wavelength is plotted, for each of the two epochs. All data points are consistent within $1-\sigma$ error bars, indicating that there is no change in transit depth with color. Overall, Figures~4 and Figure~5 lead us to the conclusion that there are no asymmetries in the observed transits of CoRoT-29b, at any wavelength interval, that can be associated to gravity darkening or other effects.

In conclusion, we observe two transits of CoRoT-29b, with a precision of around 300 ppm in the white light curve and about 400-700 ppm in several color passbands covering the optical range from 515-915 nm. We find no signs of an anomalous or asymmetric light curve as reported in previous studies. We note that the transits reported by \citetads{2015A&A...579A..36C} were taken by the CoRoT mission in summer 2011. \citetads{2015A&A...579A..36C} report also two ground-based light-curves, one taken with the 2-m Faulkes Telescope North (FTN) on 28 May 2012, and one taken on 14 July 2014 with the IAC80 telescope. Both CoRoT and the FTN data showed transits with clear and consistent asymmetries. In the IAC80 data, - taken only 2 weeks before our first GTC run - the relatively large noise permits however only a marginal identification of the asymmetry, if it was present at all. While it is not impossible that both the CoRoT and the FTN data suffered from unfortunate instrumental effects causing a similar asymmetry, another explanation is that the poorly understood source of this asymmetry disappeared sometimes between 2012 and 2014.

The Kepler mission has recently shown examples of uncommon light curves, with the detection of a possible  planetary debris disk around a white dwarf \citepads{2015Natur.526..546V}, or the puzzling behavior of the light curve of KIC 8462852 \citepads{2015arXiv150903622B}. In the near future, with the launch of missionw such as TESS and PLATO, which will cover very large fractions of the sky, the number of exoplanet discoveries around very bright stars will increase. As the total number of planets increases, the number of planets which may present challenging or unexpected light curves will also increase. Some might be first confirmations of theoretically predicted phenomena, but some may turn out to be systematic noises due to the low signal to noise of the observations or remain difficult to explain. In all cases, we demonstrated here that, using spectro-photometric techniques and having adequate instrumentation, large aperture ground-based telescopes can greatly contribute to the study of these phenomena.


\begin{acknowledgements}
Based on observations made with the Gran Telescopio Canarias (GTC), installed in the Spanish Observatorio del Roque de los Muchachos of the Instituto de Astrofisica de Canarias, in the island of La Palma. This work is partly financed by the Spanish Ministry of Economics and Competitiveness through projects ESP2013-48391-C4-2-R, ESP2014-57495-C2-1-R, and AYA2012-39346-C02-02 . G.C. also acknowledges the support by the Natural Science Foundation of Jiangsu Province (Grant No. BK20151051) and the National Natural Science Foundation of China (Grant No. 11503088). R.A. acknowledges the Spanish Ministry of Economy and Competitiveness (MINECO) for the financial support under the Ramón y Cajal program RYC-2010-06519. F.M. acknowledges the support of the French Agence Nationale de la Recherche (ANR), under the program ANR-12-BS05-0012 Exo-atmos. SH acknowledges financial support from the Spanish Ministry of Economy and Competitiveness (MINECO) under the 2011 Severo Ochoa Program MINECO SEV-2011-0187.

\end{acknowledgements}


\bibliographystyle{aa}
\bibliography{biblio_corot29}




%
%
%

%

\end{document}